\documentclass[sigplan,10pt]{acmart}
\acmConference[PPoPP'26]{ACM SIGPLAN Annual Symposium on Principles and Practice of Parallel Programming}{January 31--February 04, 2026}{Sydney, Australia}
\acmYear{2026}
\acmISBN{} % \acmISBN{978-x-xxxx-xxxx-x/YY/MM}
\acmDOI{} % \acmDOI{10.1145/nnnnnnn.nnnnnnn}
\startPage{1}

\setcopyright{none}
\usepackage{booktabs}   %% For formal tables:
\usepackage{subcaption} %% For complex figures with subfigures/subcaptions
\usepackage{pifont}

\usepackage{makecell} 
\usepackage{multirow}

\usepackage[linesnumbered,ruled,vlined]{algorithm2e}

\usepackage{tikz}

\usepackage{mathtools}
\DeclarePairedDelimiter\ceil{\lceil}{\rceil}
\DeclarePairedDelimiter\floor{\lfloor}{\rfloor}

\newenvironment{smaller_equation}{\begin{normalsize}\begin{equation*}}{\end{equation*}\end{normalsize}}

\begin{document}

%% Title information
% \title[PPoPP Paper]{PPoPP Sample Paper}
\title[SPIDER]{SPIDER: Unleashing Sparse Tensor Cores for Stencil Computation via Strided Swapping}
%% [Short Title] is optional;
                                        %% when present, will be used in
                                        %% header instead of Full Title.
%\titlenote{For simple use only}             %% \titlenote is optional;
                                        %% can be repeated if necessary;
                                        %% contents suppressed with 'anonymous'
%\subtitle{Check the format}                     %% \subtitle is optional
%\subtitlenote{Everything should just work}       %% \subtitlenote is optional;
                                        %% can be repeated if necessary;
                                        %% contents suppressed with 'anonymous'

%% Author information
%% Contents and number of authors suppressed with 'anonymous'.
%% Each author should be introduced by \author, followed by
%% \authornote (optional), \orcid (optional), \affiliation, and
%% \email.
%% An author may have multiple affiliations and/or emails; repeat the
%% appropriate command.
%% Many elements are not rendered, but should be provided for metadata
%% extraction tools.

\author{Qiqi Gu}
% \authornotemark[1]
\authornote{Equal Contribution}
\email{qiqi.gu@sjtu.edu.cn}
\affiliation{%
  \institution{Shanghai Jiao Tong University}
  \state{Shanghai}
  \country{China}
}

\author{Chenpeng Wu}
\authornotemark[1]
\email{cpwu_sjtu@sjtu.edu.cn}
\affiliation{%
  \institution{Shanghai Jiao Tong University}
  \state{Shanghai}
  \country{China}
}

\author{Heng Shi}
\email{heng.shi@sjtu.edu.cn}
\authornotemark[2]
\affiliation{%
  \institution{Shanghai Enflame Technology Co.Ltd; Shanghai Jiao Tong University}
  \state{Shanghai}
  \country{China}
}

\author{Jianguo Yao}
\email{jianguo.yao@sjtu.edu.cn}
% \authornotemark[2]
\authornote{Corresponding author}
\affiliation{%
  \institution{Shanghai Jiao Tong University}
  \state{Shanghai}
  \country{China}
}

%% Remove these two lines for the camera-ready version
\fancyhead{}  % clears the header fields that trigger HotCRP format checker
\renewcommand\footnotetextcopyrightpermission[1]{} % removes the ACM copyright footnote

%% Abstract
%% Note: \begin{abstract}...\end{abstract} environment must come
%% before \maketitle command
\begin{abstract}

Recent research has focused on accelerating stencil computations by exploiting emerging hardware like Tensor Cores. To leverage these accelerators, the stencil operation must be transformed to matrix multiplications. However, this transformation introduces undesired sparsity into the kernel matrix, leading to significant redundant computation.

In this paper, we present SPIDER, the first system to turn this unresolved sparsity into an optimization opportunity by exploring the potential of Sparse Tensor Cores (SpTCs) for stencil acceleration. Specifically, SPIDER introduces an efficient and elegant transformation method that integrates two cooperative techniques: an ahead-of-time strided swapping transformation for kernel matrices and an on-the-fly row-swapping mechanism for inputs. This rule-based approach effectively transforms stencil computation into operations compatible with SpTCs, introducing only slight compile-time overhead and zero runtime overhead. Additionally, SPIDER incorporates multiple optimizations to maximize computational efficiency.
Experimental evaluations demonstrate that SPIDER outperforms vendor library cuDNN by 6.20$\times$ and state-of-the-art (SOTA) Tensor Core-based approaches (ConvStencil, FlashFFTStencil, etc.) by 2.00$\times$ on average.

\end{abstract}

%% \maketitle
%% Note: \maketitle command must come after title commands, author
%% commands, abstract environment, Computing Classification System
%% environment and commands, and keywords command.
\maketitle

\section{Introduction}

Stencil computation, characterized as one of the seven most critical numerical methods in science and engineering\cite{pcolella,asanovic2006landscape}, is widely used across domains, including fluid dynamics\cite{huynh2014high,lusher2021opensbli}, earth modeling\cite{jacquelin2022scalable}, weather simulations\cite{ao201726,ben2022productive}, and wave equation\cite{akbudak2020asynchronous,qu2023exploiting}. It updates the points inside a grid (also known as \textit{stencil input}) by applying a predefined pattern of weighted contributions from their neighboring points (also known as \textit{stencil kernel}). Specifically, the updated value at a given location is computed as a linear combination of its adjacent elements (also known as \textit{HALO region}), where each neighbor's contribution is scaled by a corresponding coefficient. This operation is iteratively applied across the computational domain until convergence is achieved, with values at each time step derived from the results of the preceding time step.

The independent nature of point updates makes stencil computation particularly well-suited for Single Program Multiple Data (SPMD) parallel programs, which are highly effective on multi-core CPUs and GPU architectures.
Within this problem domain, researchers have pursued various optimization techniques for decades, primarily focusing on vectorization\cite{rawat2018associative, zhao2019exploiting} and tiling\cite{wolfe1989more, strzodka2010cache, bondhugula2016diamond} schemes. 
Vectorization exploits Single Instruction Multiple Data (SIMD) units in modern processors, mapping stencil computations onto vector instructions (e.g., AVX-512) for parallel execution. Building upon this, techniques such as loop unrolling\cite{deitz2001eliminating, rawat2018associative} and data layout transformations\cite{henretty2011data, henretty2013stencil} are proposed for further acceleration.
Tiling, also called blocking, partitions nested loops into smaller chunks to improve data locality. Representative tiling strategies developed specifically for stencil computation include rectangle tiling\cite{rivera2000tiling, nguyen20103}, cache oblivious tiling\cite{frigo2005cache, strzodka2010cache}, and tessellating tiling\cite{yuan2017tessellating}.

The recent emergence of specialized arithmetic logic units (ALUs) has opened new avenues for stencil computation acceleration. Notably, \textit{Tensor Cores}, specifically designed to accelerate matrix multiplication operations, are highly promising, delivering up to 4$\times$ higher throughput compared to traditional CUDA Cores. This architectural advancement has inspired a new research direction that reformulates stencil computation into a similar form as matrix multiplication to utilize Tensor Cores for acceleration. Recent studies\cite{TCStencil, chen2024convstencil, zhang2024lorastencil} have made significant progress following this idea. 
The pioneering work TCStencil\cite{TCStencil} has paved the way by transforming the stencil computation into matrix multiplication to align with the requirements of Tensor Cores. 
Although its transformation is naive and not fully optimized, TCStencil still achieves notable performance improvements.
Subsequent innovations, including ConvStencil\cite{chen2024convstencil} and LoRAStencil\cite{zhang2024lorastencil}, further advanced this paradigm through optimized memory access patterns.

However, a fundamental limitation of all existing Tensor Core-based approaches is their reliance on zero-padding during transformation. This practice inevitably introduces significant sparsity, forcing the hardware to process numerous zero-valued elements throughout the computational pipeline, wasting resources on redundant computations.

Recently, a significant evolutionary upgrade has been launched in Tensor Cores, featuring dedicated support for sparse computation. This further increases the theoretical performance ceiling by up to 2$\times$\cite{amperewhitepaper}. While this new ALU design can skip unnecessary calculations, which makes it ideally suited to address the unresolved side-effects in prior work, enabling SpTC for acceleration remains extremely challenging. 
The core difficulty lies in the hardware's rigid input constraint: it only processes data that conforms to a specific 2:4 structured sparsity pattern. 
Consequently, no existing research has succeeded in leveraging SpTC for stencil computation.

To fill the research gap and utilize the underlying sparsity, this paper presents SPIDER, a stencil computation system that transforms stencil computation into sparse matrix multiplication (SpMM) and accelerates it using SpTCs.
To summarize, the contribution of this paper is as follows:

\begin{itemize}
    \item To the best of our knowledge, this is the first work to unleash SpTCs for stencil computation. To achieve this, we propose a novel transformation strategy that converts stencil computation into SpTC-compatible form. This design turns the unresolved sparsity in existing approaches into an optimization opportunity.

    \item To adapt stencil computation for SpTCs while preserving mathematical equivalence, we propose a novel transformation integrating two cooperative techniques: an ahead-of-time strided swapping approach applied to kernel matrices and a runtime row swapping method for processing on-the-fly inputs. This joint approach is designed in rule-based manner without incurring any runtime overhead.
    
    \item The proposed transformation leverages a rule-based method, significantly simplifying its implementation. This design facilitates seamless integration into existing compiler frameworks (e.g., LLVM) and scientific computing libraries. Consequently, users can rapidly deploy the optimization via compiler passes or lightweight plug-ins, minimizing engineering overhead.
    
    \item We design and apply multiple performance optimizations including hierarchical tiling and data packing to maximize computational efficiency. Experimental results show our design outperforms vendor library cuDNN by 6.20$\times$ and SOTA Tensor Core-based approaches by 2.00$\times$ on average.
\end{itemize}

\section{Background and Motivation}

\subsection{Sparse Tensor Core}

\begin{figure}[t]
  \includegraphics[width=.95\linewidth]{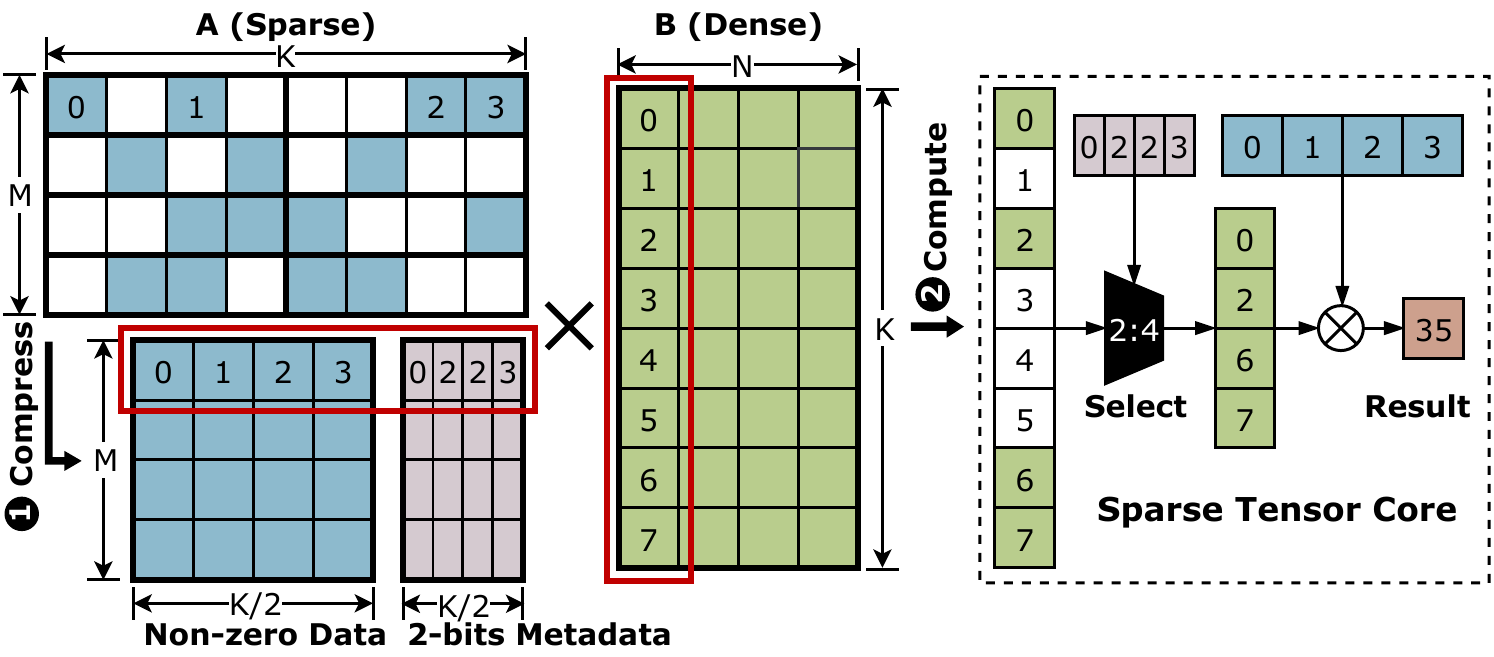} 
  \caption{2:4 Structured Sparse Format and its Compressed Representation for Sparse Tensor Cores.}
  \label{fig.sptc}
\end{figure}

Sparse ALU is an arithmetic logic unit that enhances efficiency in computing by skipping redundant operations on sparse data.
As a representative instance, Sparse Tensor Core (SpTC) was introduced by NVIDIA since the Ampere architecture in 2021\cite{amperewhitepaper}.
This unit is specifically designed to multiply a 2:4 structured sparse matrix by a dense matrix, utilizing auxiliary metadata that indicates the positions of non-zero values within the sparse matrix.
As illustrated in Figure \ref{fig.sptc}, the SpTC reuses existing Tensor Core units by incorporating an additional selection stage prior to the actual Multiply-Accumulate (MAC) computation. Guided by the metadata, this logic selects the corresponding valid values (2 out of 4 elements) from the dense matrix before applying the MAC operation with the sparse matrix. Consequently, the SpTC reduces computational workload by half, achieving up to 2$\times$ speedup compared to conventional Tensor Core units.

To utilize SpTC, the sparse matrix must first be encoded into a compressed format. This format consists of a data matrix containing the non-zero elements preserved in their original order, and a metadata matrix encoded with 2-bit descriptors. Programmers can invoke SpTC by calling the \textit{mma.sp} instruction defined in Parallel Thread Execution (PTX) Instruction Set Architecture (ISA) provided by NVIDIA\cite{ptxisa}.

\subsection{Stencil Computation} \label{sec.stencil_optimization}

Stencil computation can be characterized by three aspects: stencil shape type, dimensionality $d$, and radius $r$ (also referred to as \textit{order}). Stencil shapes are broadly categorized into two types: star stencils, which depend on the points along each axis; and box stencils, which depend on all points inside the square or cube centered around the central point. The dimensionality parameter $d$ denotes the spatial domain of stencil operation (1D, 2D, or 3D), while the radius $r$ determines the spatial dependency range. For instance, a \textbf{Box-2D2R} stencil problem features a dependent point square of spatial dimensionality $d=2$ and radius parameter $r=2$, involving 25 points in total. 

\begin{figure}[t]
  \includegraphics[width=0.95\linewidth]{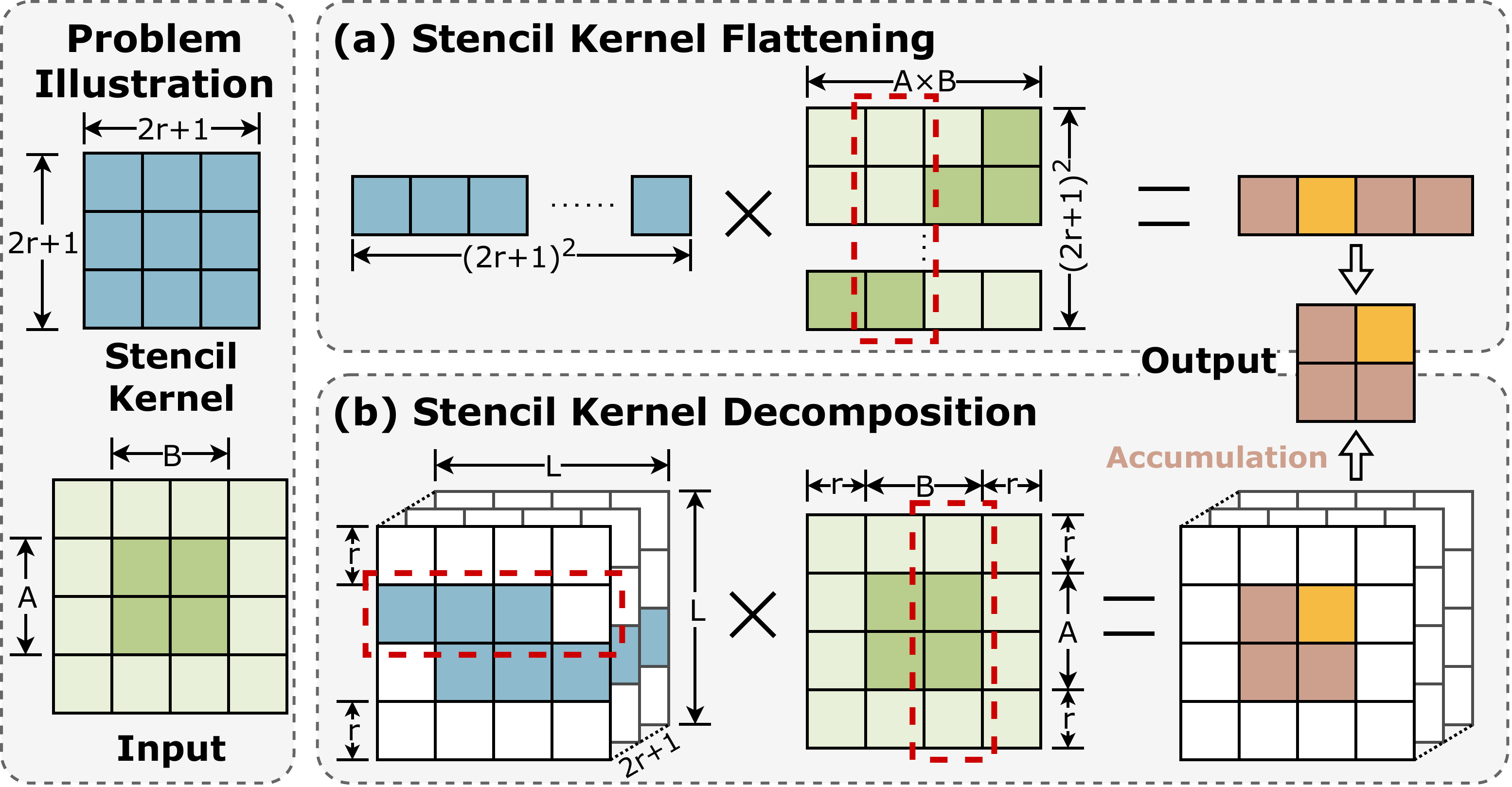} 
  \caption{Optimization Strategies for High Performance Stencil Computation.}
  \label{fig.stencil}
\end{figure}

Traditionally, stencil computation is implemented as pointwise operation, which is executed on CPUs or CUDA Cores. Recently, researchers have proposed multiple optimizations, trying to adapt stencil computation for efficient execution on Tensor Cores, further breaking the ceiling of performance. As shown in Figure \ref{fig.stencil}, these optimizations can be classified into two primary categories.

\begin{table*}\small
    \renewcommand\arraystretch{1.5}
    \tabcolsep=1.2pt
    \centering
    \begin{tabular}{cccc}
    \toprule
        Methods &  Computation Operations (FLOPS) & Input Memory Access (Elements) &  Parameter Memory Access (Elements) \\
    \midrule
        Lower Bound &  
            $AB(2r+1)^2$ & 
            $AB\frac{(c+2r)^2}{c^2}$ & 
            $AB\frac{(2r+1)^2}{c^2}$ \\
        ConvStencil &  
            $512B\ceil{\frac{A}{2c(r+1)}}\ceil{\frac{c}{8}}\ceil{\frac{r+1}{4}}\ceil{\frac{(2r+1)^2}{4}}$ ($\geq 2\textit{LB}$) & 
            $64B\ceil{\frac{(2r+1)^2}{4}}\ceil{\frac{A}{2c(r+1)}}\ceil{\frac{c}{8}}$ ($\geq 1.62\textit{LB}$)& 
            $64B \ceil{\frac{(2r+1)^2}{4}}\ceil{\frac{r+1}{4}} \ceil{\frac{A}{2c(r+1)}}\ceil{\frac{c}{8}}$ ($\geq 2.25\textit{LB}$)\\
        TCStencil   & 
            $\frac{AB}{(L-2r)^2}L^3(2r+1)$ ($\geq 4.5\textit{LB}$)& 
            $\frac{ABL^2(2r+1)}{(L-2r)^2}$ ($\geq 3\textit{LB}$)& 
            $\frac{ABL^2(2r+1)}{(L-2r)^2}$ ($\geq 3\textit{LB}$)\\
        LoRAStencil & 
            $256r\frac{AB}{c^2}\ceil{\frac{c}{8}}\ceil{\frac{2r+c}{4}}(\ceil{\frac{2r+c}{8}}+\ceil{\frac{c}{8}})$ ($\geq 1.29\textit{LB}$) & 
            $32\frac{AB}{c^2}\ceil{\frac{2r+c}{4}}\ceil{\frac{2r+c}{8}}$ ($\geq \textit{LB}$)& 
            $AB\frac{4r}{\ceil{r/4}}$ ($\geq \frac{16}{9}\textit{LB}$, for $c\geq 2$)\\
    \bottomrule
    \end{tabular}
    \caption{Redundancy Analysis of Different Methods. Lower Bound represents the theoretical optimal solution without redundancy. The inequality in parentheses represents the comparison between the cost of this method and the lower bound.}
    \label{tab:overall_redundancy}
\end{table*}

\textbf{Stencil Kernel Flattening.} 
The first optimization strategy is inspired by the conceptual similarity between stencil and convolution operations. It flattens the stencil kernel into a 1D vector and reorganizes the input data accordingly using \textit{im2col} transformation\cite{chellapilla2006high, zhou2021convolution}, converting the original problem into general matrix computation.

For instance, a basic transformation adopting this strategy is shown in Figure \ref{fig.stencil}(a). Given a Box-2D stencil, this transformation flattens the $(2r+1) \times (2r+1)$ stencil kernel into a vector of length $(2r+1)^2$, while reorganizing the $A \times B$ input data into a $(2r+1)^2 \times (A \times B)$ matrix. This transformation reformulates each stencil update as a matrix-vector multiplication (GEMV) operation. Thereby, with this transformation, Tensor Cores can be used for computation, though the overall efficiency is suboptimal.
Furthermore, ConvStencil\cite{chen2024convstencil} proposes a \textit{dual tessellation} technique that converts stencil computation into GEMM operations, thereby enabling their mapping onto Tensor Cores. Also, it eliminates memory redundancies caused by overlapping neighborhoods in the reorganized input through its \textit{stencil2row} transformation.

\textbf{Stencil Kernel Decomposing.} 
The second optimization strategy tries to enable better alignment with Tensor Core compute patterns by decomposing the stencil kernel, with each decomposed substructure contributing to a partial result. Each substructure and the corresponding input are transformed into matrix multiplication separately and the partial results are accumulated later.

For example, TCStencil\cite{TCStencil} provides a basic implementation, as shown in Figure \ref{fig.stencil}(b). It decomposes the original stencil kernel by row, and replicates each row of a stencil kernel $L-2r$ times inside a $L\times L$ matrix, allowing it to perform $L-2r$ simultaneous updates. Crucially, this transformation recasts the computation as GEMM operations, which can be executed efficiently on Tensor Cores.
Moreover, LoRAStencil\cite{zhang2024lorastencil} specifically optimizes the stencil computation by assuming stencil kernels are symmetric. It applies low-rank decomposition (LoRA) to decompose the original stencil kernel into the sum of $r + 1$ outer product vector pairs. Then each vector pair and its corresponding input is reconstructed into GEMM operations through its \textit{Residual Dimension Gathering} technique. 

\begin{figure}[t]
  \includegraphics[width=0.9\linewidth]{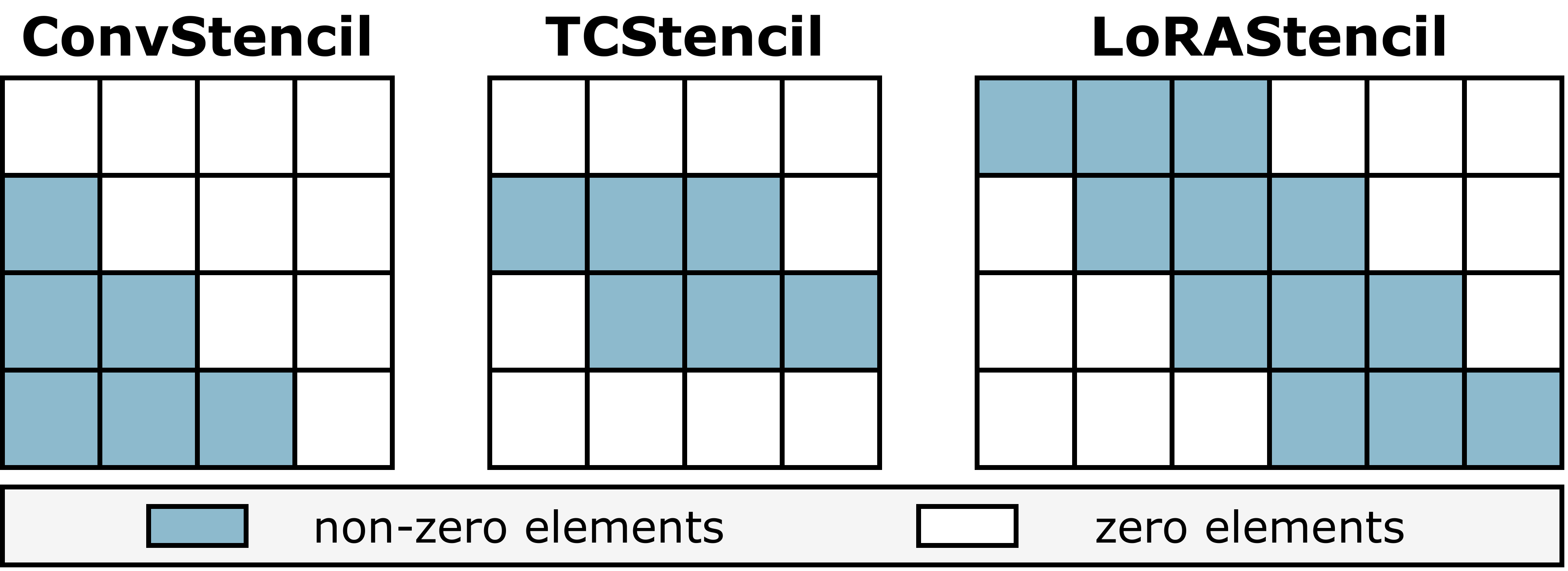} 
  \caption{Transformed Sparse Matrices in SOTA works.}
  \label{fig.sparse_matrix}
\end{figure}

Figure \ref{fig.sparse_matrix} shows the transformed kernel matrices in the aforementioned works. ConvStencil constructs kernel matrices that are either upper or lower triangular (where over half of their elements are zeros). Similarly, TCStencil and LoRAStencil introduce significant zero-padding when generating matrices for decomposed substructures. 

\textit{\textbf{Observation}: All existing works employ a common technique (zero-padding) in their transformation design, which inherently introduces sparsity into the resultant matrices.}

\subsection{Opportunity Arised from Inherent Sparsity} \label{sec:opportunity}

As discussed in aforementioned observation, existing works enforce sparsity in the transformed matrix, while their implementations still rely on dense computation paradigms on top of Tensor Cores. This discrepancy causes substantial redundancy in both computation and data movement for zero-valued elements. 

To assess the impact of this discrepancy, we formulate the computation and memory cost, quantifying the redundancy in these approaches. This formulation discusses a scenario when applying a Box-2D stencil of radius $r$ to an $A \times B$ input. For computation, we directly model the required MAC operation count per approach. Regarding memory access, we estimate the total memory access volume separately for input data and parameters. However, as this metric is strongly influenced by data reuse patterns, we normalize it by updating $c \times c$ points per tile, ensuring fair comparison across all methods. Moreover, we also formulate a theoretical lower bound, which reflects the minimal cost without zero-padding, for comparison. 

The formulations are shown in Table \ref{tab:overall_redundancy}. All of these methods exhibit significantly higher overhead compared to the lower bound. Taking a Box-2D3R stencil computation updating $8\times 8$ points per tile as an example, the computational workloads of ConvStencil, LoRAStencil, and TCStencil are approximately $2.12\times$, $2.94\times$, and $5.85\times$ higher than the lower bound, respectively. Similarly, input memory accesses exceed the lower bound by factors of $4.24\times$, $1.31\times$ and $5.85\times$ for ConvStencil, LoRAStencil, and TCStencil. Corresponding parameter memory accesses are $16.98\times$, $15.67\times$ and $23.41\times$ higher.

Notably, the relative factor for TCStencil is calculated under the configuration of updating 100 points per tile rather $8 \times 8$ points due to its design limitation. This particular configuration benefits TCStencil in this comparison, as updating more points per tile improves data reuse.
Moreover, LoRAStencil exhibits lower parameter access than the theoretical lower bound when $c=1$. This divergence arises because LoRAStencil explicitly leverages symmetric kernels within its optimization framework, whereas our lower bound derivation is established on a more general assumption, without imposing additional requirements.

This quantification demonstrates the inefficiency caused by the discrepancy discussed above, necessitating a new computation paradigm to minimize redundancies and obtain performance gains in stencil computation. Fortunately, SpTC provides an opportunity to apply sparse computation paradigm, as it specifically targets sparse matrix multiplication, which can further accelerate stencil computation.

\subsection{Challenges} \label{sec.challenge}

Despite the substantial potential of leveraging SpTC hardware to accelerate stencil computation, the transformation of such computation pattern into an SpTC-compatible form remains highly non-trivial, posing significant barriers to its practical adoption. In this section, we analyze the primary challenges encountered in designing the transformation.

\subsubsection{Alignment with Hardware Specification} 

The SpTC hardware accelerates matrix multiplication by mandating that its left-hand side (LHS) operand strictly adheres to a 2:4 structured sparsity pattern. This imposes two constraints on the corresponding transformed kernel matrix within stencil computation.
First, the matrix must exhibit a sparsity ratio of exactly 50\% or slightly above. Matrices with sparsity below this threshold are inherently incompatible with the SpTC hardware. Matrices excessively exceeding 50\% sparsity, however, prevent the ALU from fully exploiting sparsity, introducing computational overhead for zero-valued elements.
Second, the matrix is subject to a stricter constraint regarding its sparse pattern: every contiguous group of four elements must contain at least two zero values. Although prior work demonstrates that kernel matrix inherently exhibits sparsity, its native pattern does not satisfy SpTC requirements. Consequently, explicit structural transformation is necessary to make it compatible with the hardware specification.

\subsubsection{Maintenance of Mathematical Equivalence}

To enable SpTC in computation, the data pattern must be SpTC-compatible. To comply with these requirements, existing applications, such as Large Language Models (LLMs), explicitly construct such sparse patterns ahead of inference computation. This process typically employs training or post-training methods (e.g., parameter masking\cite{evci2020rigging, zhou2021learning, lu2023step} or model pruning\cite{sun2021dominosearch, kao2022training}). However, these approaches are fundamentally inapplicable to scientific workloads due to their strict requirements for mathematical equivalence.
To the best of our knowledge, none of the existing works have proposed a method for transforming stencil problems into SpTC-compatible formats while guaranteeing mathematical equivalence.

\section{System Design}

To perform stencil computation on SpTCs, we first introduce a novel \textit{Strided Swapping Transformation} in \S\ref{sec.transformation}, converting the computation into structured sparse matrix multiplication, conforming to SpTC specifications. Further, we propose a \textit{Zero-Cost Runtime Row-Swapping} technique in \S\ref{sec.row_swap}. Collectively, these techniques strictly guarantee mathematical equivalence between the original stencil computation and the transformed sparse matrix multiplication. Additionally, we design specific tiling and data packing optimizations to boost the computing performance in \S\ref{sec.kernel_optimization}.

\subsection{Strided Swapping Transformation} \label{sec.transformation}

\subsubsection{Converting Stencil Computation into GEMM with a Targeted Sparse Ratio} \label{sec.sparse_ratio}

To leverage the SpTC hardware, we first propose a way to transform the stencil computation into matrix multiplication operations. Following the decomposition strategy described in \S\ref{sec.stencil_optimization}, we decompose the stencil kernel into rows, and construct a separate matrix for each row, thus forming multiple GEMM operations.

\begin{figure}[t]
  \includegraphics[width=.9\linewidth]{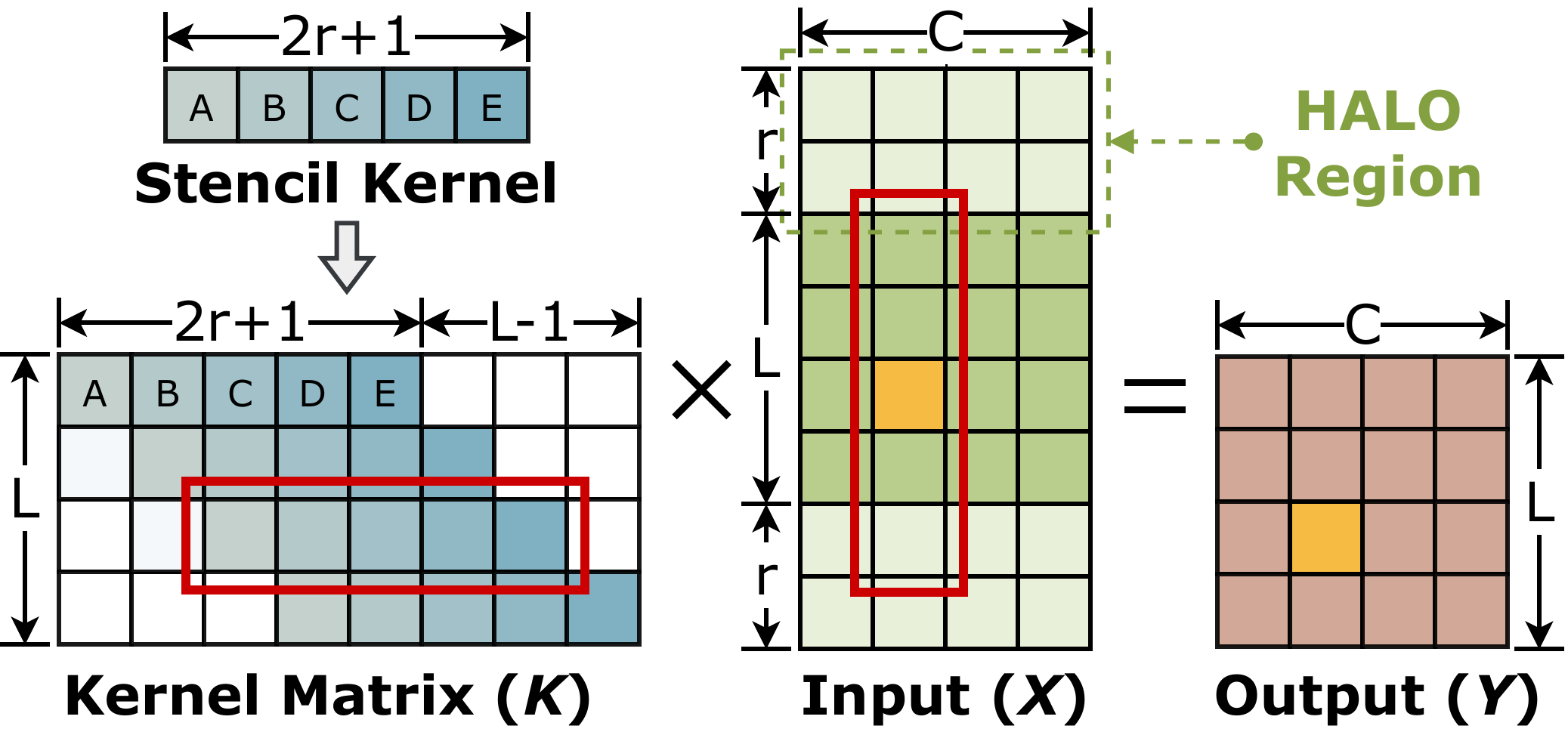} 
  \caption{Matrix Multiplication with Transformed Kernel Matrix for Stencil Computation.}
  \label{fig.gemm.design}
\end{figure}

As illustrated in Figure \ref{fig.gemm.design}, given a stencil kernel row with radius $r$, we convert it into a kernel matrix with size $L \times (2r+L)$ by repeating it $L$ times along the diagonal direction within the matrix. And the stencil computation updating $L \times C$ points is then reformulated as: $$Y = K \cdot X$$ 
where $K\in \mathbb{R}^{L \times (2r+L)}$ is the the transformed kernel matrix, $X \in \mathbb{R}^{(2r+L) \times C}$ is the input matrix containing the updated points and their $r$-radius neighborhoods, and $Y \in \mathbb{R}^{L \times C}$ is the output matrix.

The result of this matrix multiplication is strictly equivalent to the original stencil computation. For example, when computing the yellow-highlighted output element in Figure \ref{fig.gemm.design}, the original stencil computation requires multiplying the relevant input points by their corresponding coefficients (outlined in the red box). This process exactly corresponds to the behavior of matrix multiplication.

To utilize SpTC for processing the sparse transformed kernel matrix, a prerequisite is that the matrix's sparsity ratio meets the hardware requirement, specifically, at least 50\%. The sparsity ratio of the transformed kernel matrix can be formulated as:
\begin{smaller_equation} \label{formula.sparsity}
\text{Sparsity} = \frac{L \times (2r+1)}{L \times (2r+L)} = \frac{2r+1}{2r+L}
\end{smaller_equation}

This formulation indicates that variable $L$ must satisfy $L\geq2r+2$ to meet the requirement of SpTC. However, sparsity ratios significantly above 50\% are also undesirable, as SpTC can only exploit sparsity at 50\%, and higher sparsity does not yield additional benefits. Therefore, we set $L=2r+2$ to satisfy the sparsity ratio requirement while maximizing hardware utilization.

\subsubsection{Aligning with Structured Sparse Requirement of SpTC via Strided Swapping} \label{sec.sparse_pattern}

\begin{figure}[t]
  \includegraphics[width=.9\linewidth]{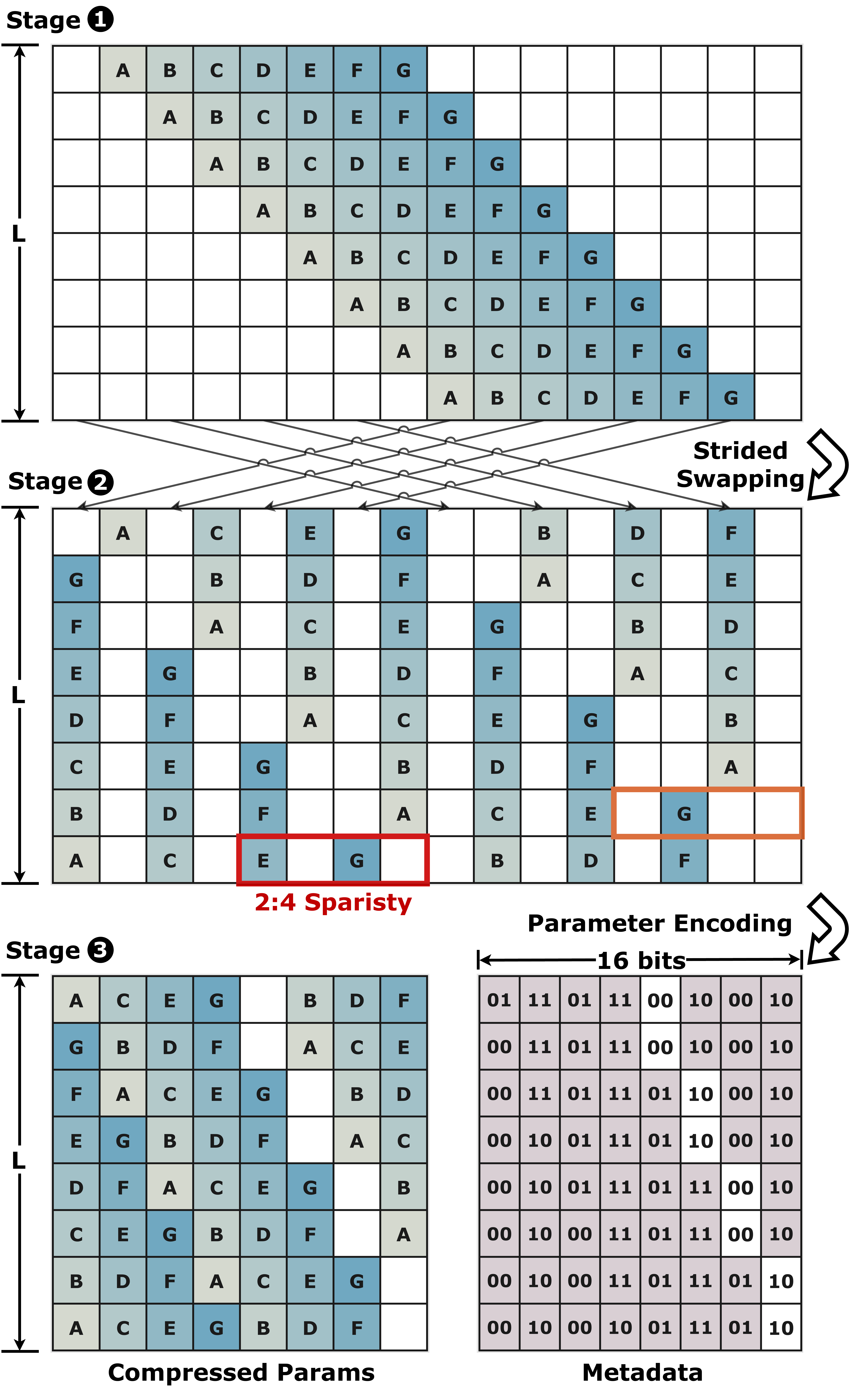} 
  \caption{Strided Swapping and Encoding of Kernel Matrix. As described in \S\ref{sec.sparse_ratio}, the kernel matrix is configured with $L=2r+2$. In this illustration, we set radius $r=3$.}
  \label{fig.sparsify}
\end{figure}

Beyond the sparsity ratio requirement, SpTC additionally requires that the transformed matrix adhere to a designated sparsity pattern, specifically 2:4 structured sparsity. This section therefore introduces a strided swapping method and a corresponding encoding phase to meet the hardware specification of SpTC, as shown in Figure \ref{fig.sparsify}.

Stage \ding{202} in Figure \ref{fig.sparsify} shows the initial state of this swapping method, which is the transformed kernel matrix generated in \S\ref{sec.sparse_ratio}. Notably, we pad two columns to this matrix (from $8\times 14$ to $8\times 16$), to align with the shape requirement of SpTC. In this matrix, the non-zero elements are aggregated, violating hardware structured sparsity requirements. Fortunately, this matrix exhibits a highly regular sparse pattern, where all non-zero elements are located in the central parallelogram region.
Leveraging this regularity, we propose a strided column swapping method to align with the 2:4 structured sparse pattern required by SpTC. As illustrated in Figure \ref{fig.sparsify}, we swap odd-indexed columns $j$ with columns $j+L$, while preserving even-indexed columns unchanged. The outcome of this strided swapping is shown in Stage \ding{203}, ensuring every contiguous 4-element segment contains at most two non-zero values and satisfying the 2:4 sparse pattern required by SpTC hardware.

A key advantage of this permutation is its simplicity. Although permuting the kernel matrix necessitates runtime reordering of the corresponding input data to preserve mathematical equivalence, this simplicity forms the basis for low-overhead runtime row swapping, which will be presented in \S\ref{sec.row_swap}.
Meanwhile, it should be noted that this permutation does not require the stencil kernel to follow a particular shape or numerical pattern, allowing generalization to any stencil problem.

Then, we introduce an encoding phase to further compress this structured sparse matrix into a hardware-compatible format. As shown in Stage \ding{204}, this compression produces two matrices. First, non-zero values are compressed into a value matrix. Second, positional metadata is generated, where a 2-bit encoding specifies the position of each non-zero within the segment. 
For example, the four elements \colorbox{gray!15}{\texttt{E0G0}}, outlined in the red box in Figure \ref{fig.sparsify}, are encoded as \colorbox{gray!15}{\texttt{EG}} with metadata \colorbox{gray!15}{\texttt{00 10}}. Here, metadata \colorbox{gray!15}{\texttt{00}} and \colorbox{gray!15}{\texttt{10}} denote that the data \colorbox{gray!15}{\texttt{E}} and \colorbox{gray!15}{\texttt{G}} are at the first and third position in the segment, respectively.
In addition, for certain special cases where a segment contains only one non-zero element, we retain one zero as a placeholder in the value matrix and generate corresponding metadata. For instance, for the elements highlighted in the orange box (\colorbox{gray!15}{\texttt{0G00}}), the compressed value becomes \colorbox{gray!15}{\texttt{G0}} with corresponding metadata \colorbox{gray!15}{\texttt{01 10}}.
% To maintain the consistency between the compressed values and their positional encodings, 
All metadata is stored in an increasing order, starting from the least significant bit within each segment.

The transformation rule is uniform for any stencil of a given radius $r$. This enables a predefined extraction rule and metadata for non-zero elements, simplifying implementation. Critically, this entire process can be performed offline, eliminating any runtime overhead.

\textbf{Quantitative Analysis}
Similar to previous comparison in \S\ref{sec:opportunity}, we can formulate the computational workload and memory access volume for SPIDER as follows:
\begin{smaller_equation}
    \begin{aligned}
        \text{SPIDER}_{C} &= 256\frac{AB}{c^2}(r+1)\ceil{\frac{c}{8}}^2\ceil{\frac{2r+c}{4}} \\
        \text{SPIDER}_{I} &= 32\frac{AB}{c^2}(2r+1)\ceil{\frac{c}{8}}\ceil{\frac{2r+c}{4}} \\
        \text{SPIDER}_{P} &= 16\frac{AB}{c^2}(2r+1)\ceil{\frac{c}{8}}\ceil{\frac{2r+c}{4}}
    \end{aligned}
\end{smaller_equation}
here, $\text{SPIDER}_{C}$ denotes the computation operations required by SPIDER, whereas $\text{SPIDER}_{I}$ and $\text{SPIDER}_{P}$ represent the memory accesses for input and parameter, respectively.

Supposing a Box-2D3R stencil problem updating $8\times 8$ points per tile, we can compare the computation and memory cost for updating a single point using our design against existing approaches, as illustrated in Table \ref{tab:Quantitative_Analysis}. 

\begin{table}[t]\small
    \centering
    \resizebox{0.95\linewidth}{!}{
        \begin{tabular}{cccc}
        \toprule
            Methods & \makecell[c]{Computation \\ Operations} & \makecell[c]{Input \\ Memory Access} & \makecell[c]{Parameter \\ Memory Access} \\
        \midrule            
             Lower Bound    &  49           &  3.06         & 0.77  \\
             ConvStencil    &  104          &  13           & 13    \\
             TCStencil      &  286.72       &  17.92        & 17.92 \\
             LoRAStencil    &  144          &  4            & 12    \\
             SPIDER         &  \textbf{56}  &  14           & \textbf{7} \\
        \bottomrule
        \end{tabular}
    }
    \caption{Quantitative Comparison of Computation and Memory Costs for Point Update in the Box-2D3R Stencil Problem.}
    \label{tab:Quantitative_Analysis}
\end{table}

As demonstrated, SPIDER significantly outperforms all competitors in both computational workload and parameter access efficiency, because our strided swapping transformation reduces the requirement of loading and computing zero values. Regarding input data access, our method is comparable to or better than alternative approaches, except for LoRAStencil. It should be emphasized that LoRAStencil is limited to symmetric stencil kernel configurations, which makes it incompatible with general stencil problems (unlike SPIDER).

\subsection{Zero-Cost Runtime Row Swapping}\label{sec.row_swap}

\begin{figure}[t]
  \centering 
  \begin{minipage}[b]{0.50\linewidth}
    \centering
    \includegraphics[width=\linewidth]{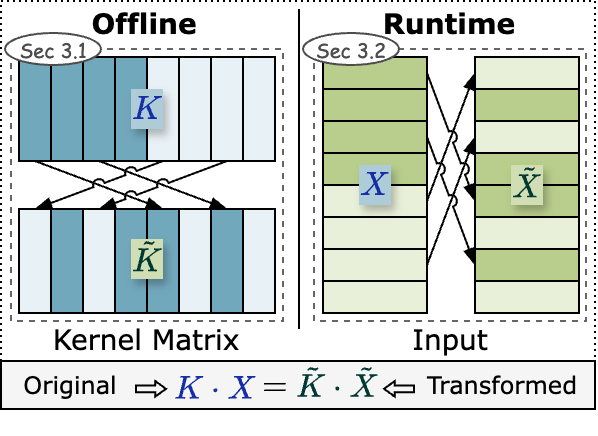}
    \caption{Corresponding Row Swap to Ensure Mathematical Equivalent.}
    \label{fig.row_swap}
  \end{minipage}
  \hfill 
  \begin{minipage}[b]{0.46\linewidth}\scriptsize
    \centering
    \begin{tabular}{c cc}
    \toprule
       \multirow{2}{*}[-0.8ex]{Metrics} & \multicolumn{2}{c}{Row Swapping} \\
            \cmidrule(lr){2-3}
            & Without & With \\
    \midrule
        \makecell[c]{Memory \\ Throughput \\ (GB/s)}& 666.14 & 665.90 \\ 
        \addlinespace[0.7ex]
        \makecell[c]{Instruction \\ Counts (K)} & 131123.2 & 131123.2 \\
        \addlinespace[0.7ex]
        \makecell[c]{Duration (µs)} & 641.95 & 641.89 \\
    \bottomrule
    \end{tabular}
    \vspace{13pt}
    \captionof{table}{Row Swapping Cost Evaluation in SPIDER.}
    \label{tab:row_swap_comparision}
  \end{minipage}
\end{figure}

To preserve mathematical equivalence, after strided swapping on the kernel matrix (\S\ref{sec.sparse_pattern}), the input matrix needs to be transformed correspondingly. To be specific, as shown in Figure \ref{fig.row_swap}, swapping columns $i$ and $i+L$ (where $i=0,2,\cdots,L-1$) in the kernel matrix mandates swapping the $i$-th row with the $i+L$-th row in the input matrix. 
Since the stencil kernel remains constant, its transformation can be performed offline. In contrast, the input is updated iteratively, requiring runtime input transformation. 
This overhead is amplified by thousands of iterations, demanding careful optimization to mitigate the associated costs.

Intuitively, we can simply implement the row swapping through explicit data copy. However, this approach inevitably introduces substantial overhead. To address this limitation, we attempt to hide the row-swap overhead during data movement from shared memory to registers, which is a standard operation in GPU computing implementations. Fortunately, the column-swap process exhibits a highly regular pattern, enabling an efficient and elegant design of corresponding row swapping.

Specifically, we combine row swapping with memory access offset calculation to implement an implicit row swapping technique.
Taking the Box-2D7R stencil as an example, the stencil computation invokes the SpTC instruction \textit{mma.sp.m16n8k16} twice. 
According to the PTX instruction specification for the RHS operand, each thread manages four elements. The thread-to-row mapping for the $i$-th element in the original workflow is defined by:
\begin{smaller_equation}
    \textit{offset}_{\textit{row}} = 2 \times (\textit{lane\_id} \bmod 4) + 8\floor{\frac{i}{2}}+ (i \bmod 2)
\end{smaller_equation}
here, $\textit{lane\_id}\in\{0,1,\cdots,31\}$ denotes the thread index within a warp.

Our row swapping strategy merely adds an extra term to this equation, the updated thread-to-row mapping is given by:
\begin{smaller_equation}
\begin{aligned}
    \tilde{\textit{offset}}_{\textit{row}} = \textit{offset}_{\textit{row}} + \begin{cases}
        16(-1)^k, & \text{if } i \bmod 2 \equiv 0\\
        0, & \text{if } i \bmod 2 \equiv 1
    \end{cases}
\end{aligned}
\end{smaller_equation}
where $k\in\{0,1\}$ indicates the SpTC invocation index.

We compare implementations with and without runtime row swapping integration and evaluate the cost brought by our proposed row swapping strategy in terms of memory efficiency and instruction count, with results shown in Table \ref{tab:row_swap_comparision}.
Firstly, the integrated row swapping does not affect data movement efficiency. Specifically, each thread continues to transfer the same data volume (four elements per thread) as before. Moreover, the memory access pattern within a warp remains unchanged, preventing the introduction of additional bank conflicts that could otherwise degrade efficiency. This is corroborated by statistically equivalent memory throughput, where the 0.04\% variance falls within experimental error margins.
Next, the additional term in offset calculation can be optimized to a constant value by compiler through loop unrolling. Consequently, this incurs no additional instructions in the generated kernel, as verified by the identical instruction counts shown in Table \ref{tab:row_swap_comparision}.

Collectively, these factors yield nearly identical execution times between both methods, with a variance of merely 0.01\%. These results conclusively demonstrate that our row swapping method achieves zero runtime cost.

\subsection{Computing Optimization for SPIDER} \label{sec.kernel_optimization}

The performance of SPIDER is specifically optimized for stencil computation patterns, adhering to classical high-performance computing optimizations including tiling and data packing.

\subsubsection{Tiling Strategy for Maximum Data Reuse} \label{sec.tiling}

Tiling is a widely adopted optimization technique that decomposes data into manageable blocks, thereby exploiting data locality to enhance computational performance\cite{van2015blis}.
While tiling has been well-established for operations like GEMM, stencil features a specialized data reuse pattern, necessitating a dedicated tiling strategy that is specifically tailored for stencil computation on SpTC.

\begin{figure}[t]
  \includegraphics[width=.98\linewidth]{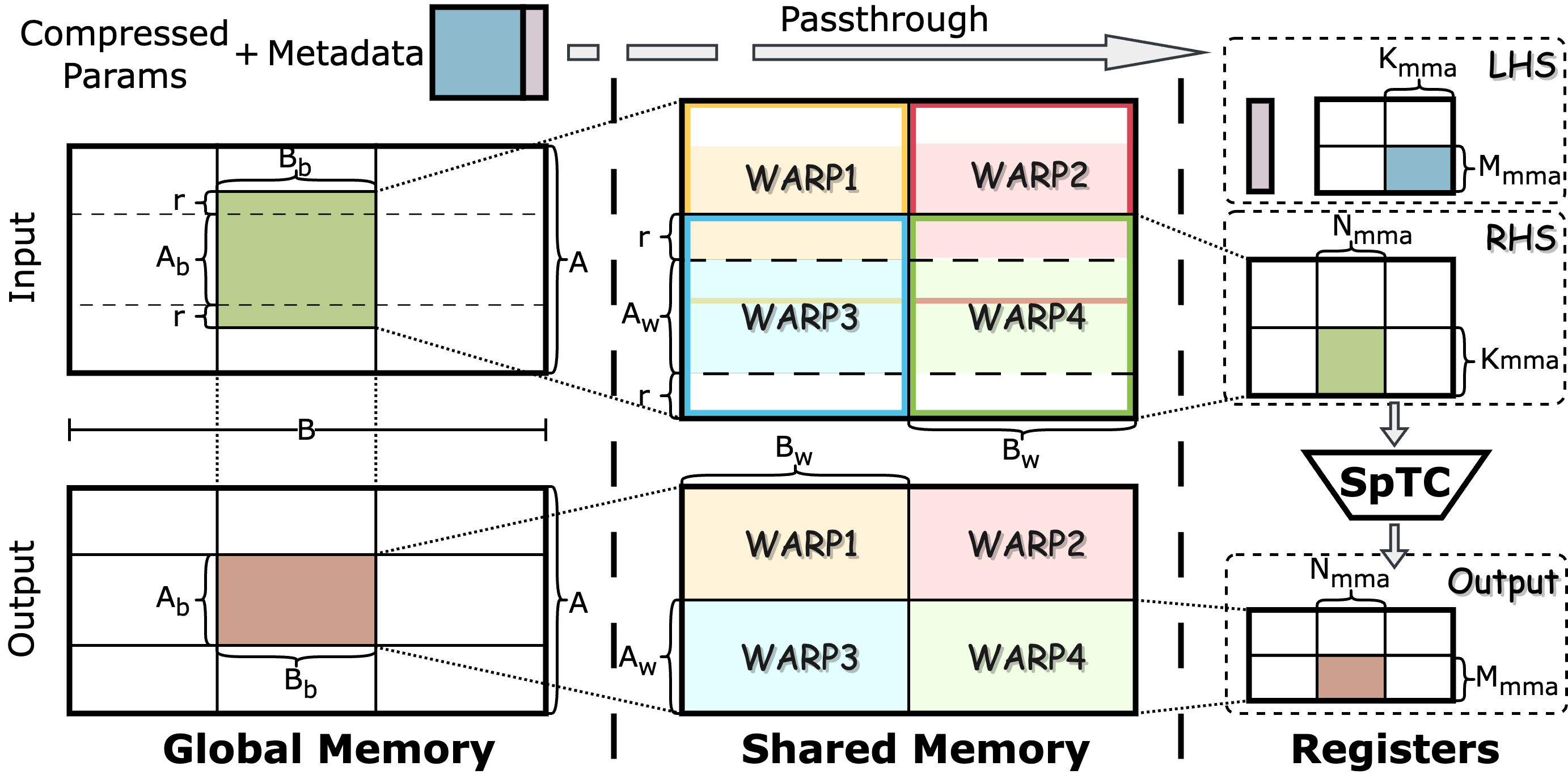} 
  \caption{Tiling Strategy for SPIDER Computing.}
  \label{fig.tiling}
\end{figure}

As illustrated in Figure \ref{fig.tiling}, SPIDER introduces a hierarchical tiling strategy across three levels of GPU memory hierarchy: (1) block-level tiling regulates the data loaded from global memory to shared memory, (2) warp-level tiling loads data from shared memory to registers for computation, and (3) mma-level tiling further partitions registers to align with the instruction shape supported by SpTC.
To be specific, consider performing a stencil operation with radius $r$ to an input of size $A \times B$. During SPIDER computing, each thread block computes an output tile of size $A_b \times B_b$, requiring loading $(A_b + 2r) \times B_b$ input (including elements in \textit{HALO region}) into shared memory. Further, the loaded data inside shared memory is partitioned into tiling size of $A_w\times B_w$, which can be scheduled for concurrent execution by a warp of 32 threads. Finally, at the hardware instruction level, the tiling size ($M_{mma}$, $N_{mma}$, $K_{mma}$) aligns with vendor-provided sparse MMA instructions. For instance, the \textit{mma.sp.m16n8k16} instruction defines dimensions $M_{mma}=16$, $N_{mma}=8$, $K_{mma}=16$.

Notably, considering the kernel matrix is reused for each tile, it resides entirely in registers throughout computation. This approach obviates the need for warp-level tiling. Consequently, the kernel matrix bypasses the shared memory hierarchy to improve overall efficiency.

\subsubsection{Data Packing for Efficient Memory Access}

\begin{figure}[t]
  \includegraphics[width=.98\linewidth]{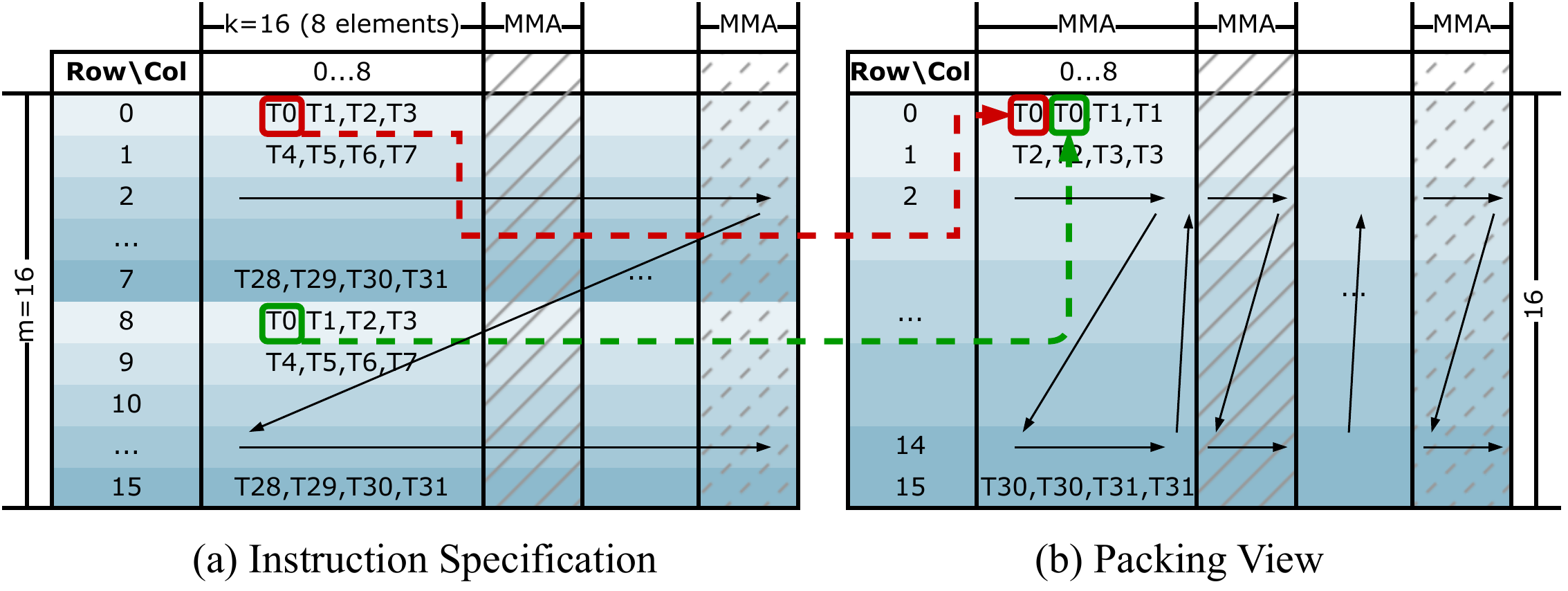} 
  \caption{Data Packing for Compressed Kernel Matrix on \textit{mma.sp.m16n8k16} Instruction.}
  \label{fig.packing_a}
\end{figure}

Modern GPUs, as highly parallel compute accelerators, often exhibit performance bottlenecks at memory access that constrain overall throughput. To mitigate this limitation, data packing has emerged as a fundamental optimization technique in GPU computing kernel design. This approach reorganizes data in memory to align with GPU-friendly memory access patterns, significantly enhancing memory throughput and computational efficiency.
In our design, the novel tiling strategy, along with specialized SpTC instructions, motivates customized data packing strategies.

For example, as illustrated in Figure \ref{fig.packing_a}(a), the fragment layout for the kernel matrix specified by SpTC introduces non-contiguous memory access pattern across threads when involving multiple MMA instructions.
Accessing these scattered memory causes uncoalesced memory access, which will increase memory transactions during data movement from global memory to registers, thereby degrading overall memory efficiency. To address this, our design proactively packs the kernel matrix into an optimized format, as depicted in Figure \ref{fig.packing_a}(b).
Specifically, data accessed by each thread is contiguously packed in memory. Furthermore, data utilized across different MMA invocations is sequentially packed.
This dual packing strategy ensures that threads within a warp access contiguous memory elements, aligning with the granularity of memory transactions.

\begin{figure}[t]
  \includegraphics[width=.98\linewidth]{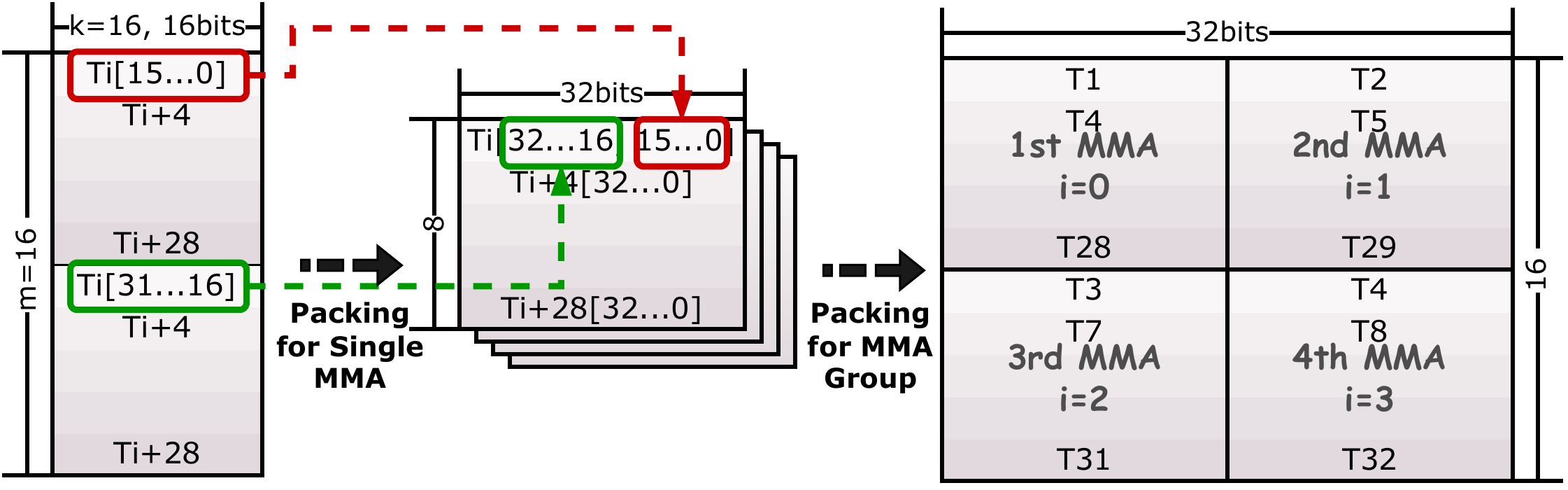} 
  \caption{Data Packing for Metadata Matrix on \textit{mma.sp.m16n8k16} Instruction.}
  \label{fig.packing_metadata}
\end{figure}

\begin{figure*}[t]
  \includegraphics[width=.98\linewidth]{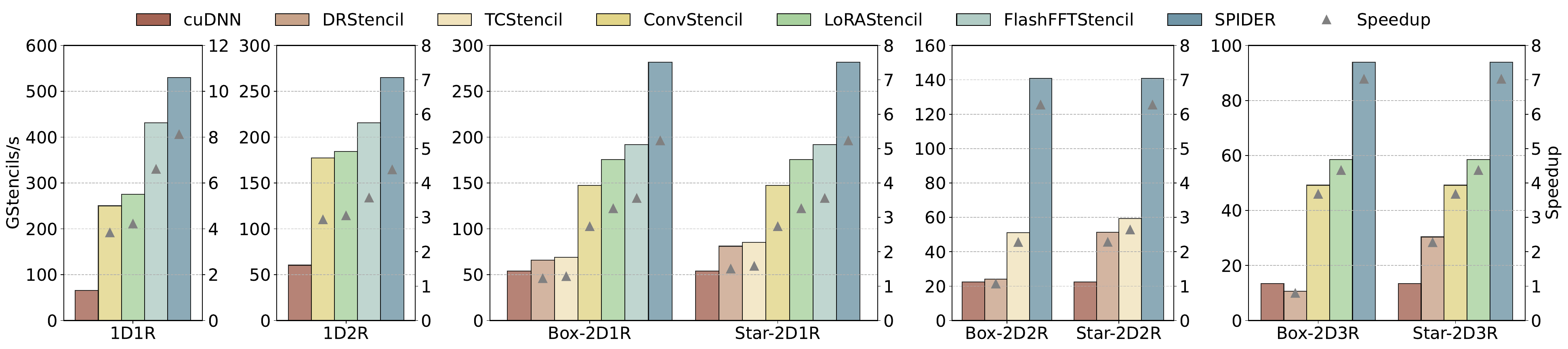} 
  \caption{Performance Comparison between SPIDER and SOTA implementations.}
  \label{fig.exp.baselineCompare}
\end{figure*}

Moreover, the metadata matrix generated by Strided Swapping Transformation (\S\ref{sec.sparse_pattern}) is also packed to enhance data reuse, as illustrated in Figure \ref{fig.packing_metadata}.
Firstly, we reorganize the metadata required by a single MMA invocation, ensuring contiguous memory access per thread, thereby reducing the number of memory requests.
Subsequently, we recognize that the SpTC specification mandates that each thread allocates a dedicated register for metadata, yet it only utilizes the metadata from eight threads. To optimize register utilization, we concatenate metadata matrices for multiple MMA instructions and specify the active set of eight threads through a sparsity selector for SpTC invocation. This approach reduces the registers allocated for each thread while preserving computational integrity.
With reduced memory requests and register allocation, the overall memory efficiency can be significantly enhanced through this metadata packing.

\section{Evaluation}

\subsection{Evaluation Setup}

\textbf{Platforms.} Our evaluation platform utilizes two Intel Xeon Platinum 8558P CPUs, 512GB DDR5 memory, and four NVIDIA A100-80GB PCIe GPUs. The GPUs are based on the Ampere architecture, which features dedicated SpTCs for sparse matrix computation. The system runs Ubuntu 22.04 LTS with CUDA 12.8 and cuDNN 9.8.0. Meanwhile, CPU frequency scaling is disabled across all experiments to ensure fairness.

\textbf{Baselines.} Our evaluation baselines encompass two categories: CUDA Core-based approaches and Tensor Core-accelerated methods. The former includes \textit{cuDNN}\cite{chetlur2014cudnn}, an  industry-standard library from NVIDIA, which provides meticulously hand-tuned implementations of primitive operators for modern GPU architectures, and \textit{DRStencil}\cite{you2021drstencil}, a SOTA framework for automatically generating and tuning highly efficient stencil code via multiple optimizations. The Tensor Core-based solutions proposed recently include \textit{TCStencil}\cite{TCStencil}, which pioneers in conducting stencil computation on Tensor Cores; \textit{ConvStencil}\cite{chen2024convstencil}, which introduces a \textit{stencil2row} transformation to minimize redundant memory accesses in general-purpose stencil operations; \textit{LoRAStencil}\cite{zhang2024lorastencil}, which leverages low-rank adaptation under symmetric kernel assumptions to further reduce memory access redundancy; and \textit{FlashFFTStencil}\cite{han2025flashfftstencil}, which employs the Fast Fourier Transform (FFT) algorithm to enhance arithmetic intensity, thereby boosting Tensor Core utilization.

It should be noted that these baseline methods are implemented using different numerical precisions: for example, ConvStencil operates in FP64 precision, whereas TCStencil employs FP16 precision. To normalize the performance, we scale the results by a factor of 4, aligning with the approach adopted in prior work\cite{chen2024convstencil, zhang2024lorastencil, han2025flashfftstencil}.

% TODO 是否保留？
\textbf{Benchmarks.} The performance of stencil computation is evaluated through benchmarks on a series of stencil shapes, including 1D stencil kernels with radius 1 and 2, and 2D stencil kernels with two common shapes (star and box), where the radius ranges from 1 to 3.

\textbf{Metrics.} To quantify the performance of the stencil operation, we adopt the widely used metric in stencil computation\cite{yuan2017tessellating, chen2019versatile, matsumura2020an5d, yuan2019tessellating, zhang2023perks, zhang2023revisiting, zhang2025jigsaw}, Stencils/s (also known as Cells/s), which represents the number of points updated per second by stencil computation. For brevity, this metric is converted to GStencils/s, where $1$ GStencils/s equals $10^9$ Stencils/s.

\subsection{Performance Comparison}

In this section, we compare the stencil computation performance of SPIDER against all baseline methods across various stencil shapes. The configurations of problem size are represented in the form of (A,B). For 1D stencil shapes, we configure the problem size to (1, 10240000). Meanwhile, 2D benchmarks employ a grid configuration of (10240, 10240).

As shown in Figure \ref{fig.exp.baselineCompare}, our approach consistently surpasses all baseline methods. Against CUDA Core-accelerated implementations, SPIDER delivers average speedups of 6.20$\times$ over cuDNN and 4.71$\times$ over DRStencil.
For Tensor Core-based solutions, it achieves average speedups of 3.13$\times$, 1.88$\times$, 1.63$\times$, and 1.35$\times$ over TCStencil, ConvStencil, LoRAStencil, and FlashFFTStencil, respectively. 
SPIDER demonstrates higher throughput than existing SOTA approaches, whether optimized for memory access (e.g., ConvStencil, LoRAStencil) or computation (e.g., FlashFFTStencil). 
This performance advantage demonstrates the effectiveness and efficiency of SPIDER to resolve the inherent sparsity in stencil computation identified in §\ref{sec:opportunity}.

Furthermore, our results demonstrate that SPIDER exhibits strong generalizability across all stencil shapes. First, it maintains stable performance across both box-shaped and star-shaped stencils. Unlike baselines such as DRStencil and TCStencil, which achieve higher performance on star-shaped patterns through specialized optimizations, our approach employs an optimization strategy targeting box shapes that can be generalized to any stencil shapes. Second, SPIDER scales efficiently with increasing stencil order. For instance, its speedup over DRStencil increases from 4.27$\times$ (Box-2D1R) to 8.82$\times$ (Box-2D3R). This advantage arises because larger radius expands the tuning search space for DRStencil, leading to suboptimal auto-tuned implementation under fixed time budget. SPIDER avoids this limitation through a predefined optimization strategy that eliminates costly search overhead.

It should be noted that our evaluation does not reflect the offline transformation overhead present in some baselines.
For instance, DRStencil heavily depends on its tuning process with a configurable time budget, which is set to 1 hour in our experiments. And this tuning process is needed once the stencil kernel shape or input size changes, resulting in significant overhead beyond actual execution. Moreover, some baselines need comprehensive transformations when adapting to Tensor Cores. Specifically, FlashFFTStencil relies on FFT computation with $O(L^2\log L)$ computational complexity and LoRAStencil employs low-rank adaptation with $O(L^3)$ computational complexity, incurring substantial computational overhead.
In contrast, SPIDER incurs minimal offline transformation overhead with $O(1)$ computational complexity, as it only reorganizes stencil kernel purely guided by pre-defined rules. 
This efficiency further highlights SPIDER's practical advantage over competing approaches when accounting for these critical preparation costs.

\subsection{Performance Trend Analysis} \label{sec.trend}

\begin{figure}[t]
  \includegraphics[width=0.9\linewidth]{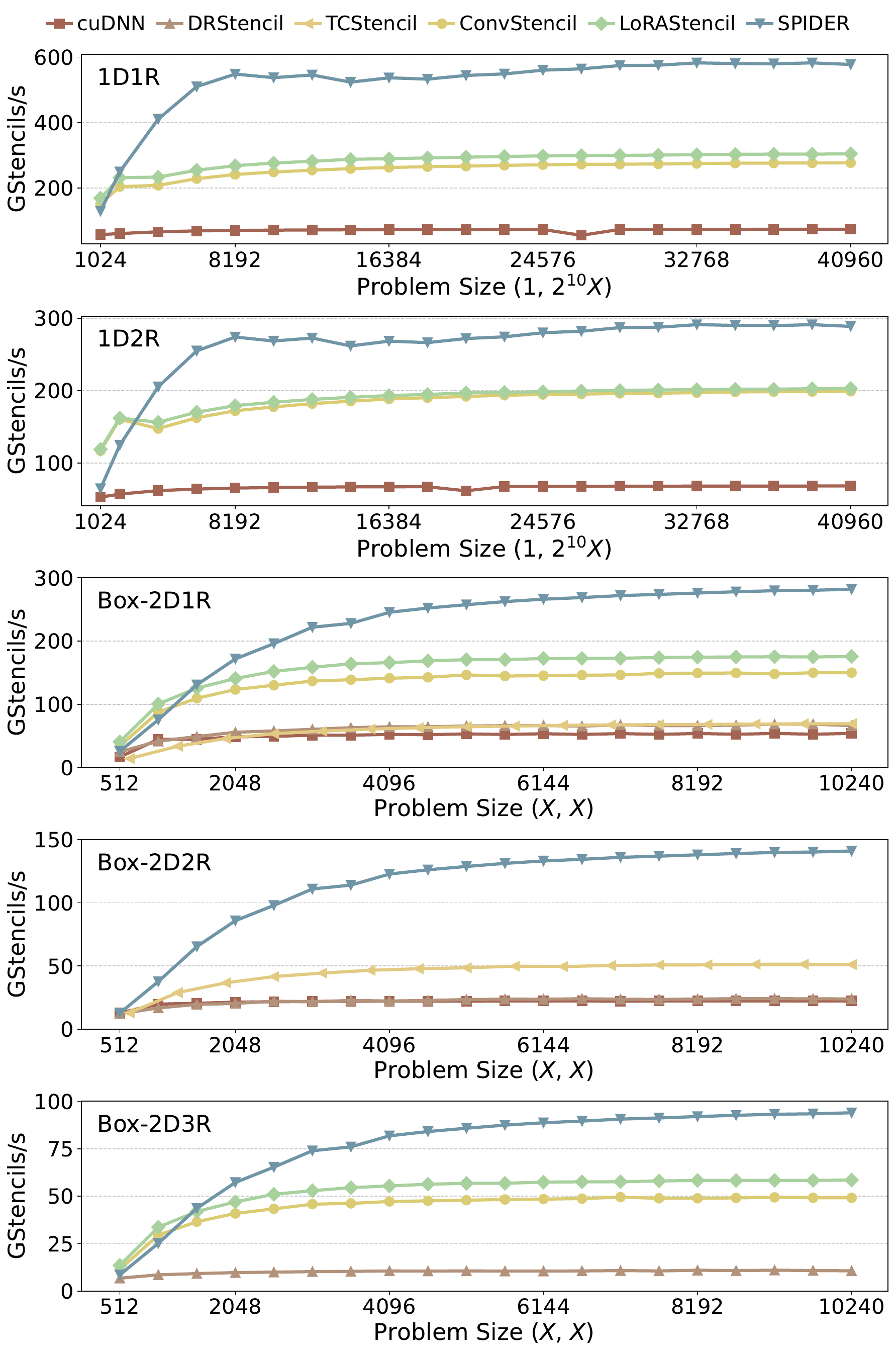}
  \caption{Performance Trend with Increasing Problem Sizes under Different Stencil Shapes.}
  \label{fig.exp.TrendAnalysis}
\end{figure}

Beyond evaluating benchmark performance on specific problem, we conduct a comprehensive analysis on the performance trend of SPIDER with increasing problem size. The configurations span problem sizes from (1, 1024$\times$256) to (1, 1024$\times$40960) for 1D cases and (512, 512) to (10240, 10240) for 2D problems. 
Here, TCStencil evaluates slightly different problem sizes due to implementation constraints, but we ensure all baselines use the same evaluated range for fairness.

As depicted in Figure \ref{fig.exp.TrendAnalysis}, SPIDER demonstrates progressive performance gains with increasing problem size until achieving peak efficiency. Upon reaching a stable performance plateau, it consistently outperforms all baseline methods, delivering 1.86$\times$ average throughput than the best-performing baseline. This scaling behavior stems from dynamic resource utilization characteristics. At smaller scales, the available GPU resource exceeds computation demands, leading to underutilization of resources (e.g., memory bandwidth, ALUs). As problem size expands, more resources are engaged during execution, enhancing GPU occupancy and thereby increasing throughput. Beyond a critical problem size, all GPU resources become saturated, limiting throughput to the peak computational capacity of the GPU, representing the theoretical performance ceiling for our approach.

Notably, SPIDER exhibits relatively lower performance compared to certain baselines (e.g., ConvStencil and LoRAStencil) at smaller problem sizes. This behavior originates from insufficient parallelism in SPIDER under these workloads, as our optimized tiling strategy employs a large tiling size for efficient memory access. 
As problem size increases, the performance of SPIDER scales with increased parallelism, thereby outperforming other baselines.
Meanwhile, beyond the performance plateau (e.g., problem size $>$(8192, 8192) for Box-2D2R), SPIDER exhibits a minor yet consistent throughput gain (up to 2\%). This can be attributed to the fixed GPU launch overhead becomes a diminishing fraction of the total execution time as workload scales.

\subsection{Ablation Study on System Optimizations}

\begin{figure}[t]
  \includegraphics[width=.98\linewidth]{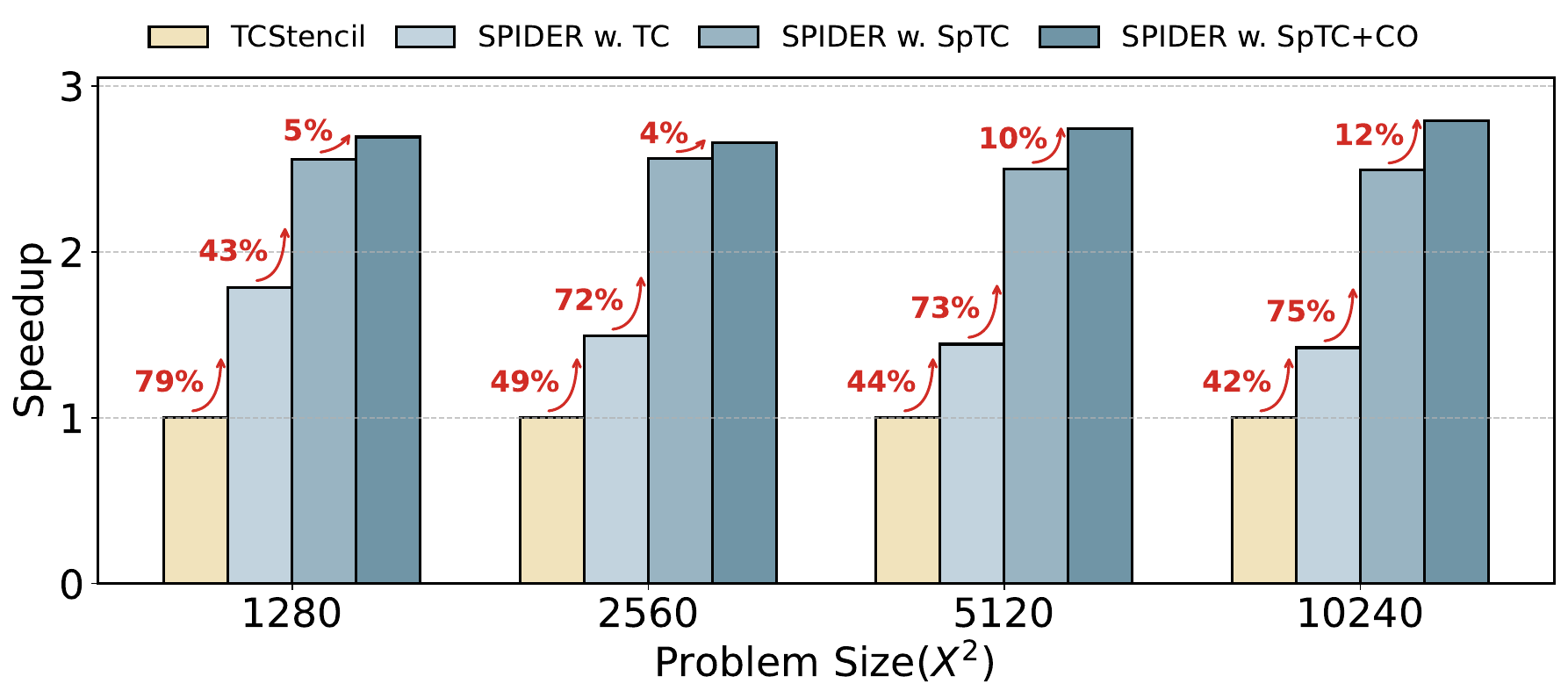}
  \caption{Performance Breakdown of SPIDER with Box-2D2R Stencil (\textit{TC}: Tensor Core, \textit{SpTC}: Sparse Tensor Core, \textit{CO}: Computing Optimization). }
  \label{fig.exp.Ablation}
\end{figure}

In this section, we breakdown the design of SPIDER and evaluate each component to quantify its performance benefits. We use TCStencil, a baseline implementation that accelerates stencil computation with Tensor Cores, as our reference point. We then incrementally apply our optimizations and measure their contributions to overall performance. Figure \ref{fig.exp.Ablation} presents the performance across different optimizations.

The first optimization, \textit{SPIDER w. TC}, which transforms stencil computation into GEMM with 50\% sparsity in the kernel matrix as described in \S\ref{sec.sparse_ratio}, achieves an average speedup of 1.54$\times$ over the baseline. This improvement stems from the elimination of redundant regions in the stencil kernel matrix compared to TCStencil, thereby increasing Tensor Core utilization.

Next, by incorporating the Strided Swapping Transformation (introduced in \S\ref{sec.sparse_pattern}), stencil computation achieves a performance boost of up to 1.75$\times$ and an average improvement of 1.66$\times$. This significant speedup naturally comes from leveraging the inherent sparsity. By utilizing SpTC for computation, nearly half of the redundant computation and memory workloads can be eliminated. This further validates our design that uses SpTCs to accelerate stencil computation.

Notably, when the problem size is (1280, 1280), the speedup achieved by incorporating SpTC is 1.43$\times$, which is lower than the average 1.74$\times$ speedup we observe in other problem sizes. The root cause of this discrepancy lies in the lower achieved occupancy of our current SpTC-incorporated implementation on small problem sizes, which leads to underutilized hardware resources, consistent with the throughput ramp-up phase analyzed in §\ref{sec.trend}.

Finally, we apply computing optimizations introduced in \S\ref{sec.kernel_optimization}, denoted as \textit{Stencil w. SpTC+CO}. These results demonstrate that our performance optimizations deliver an average speedup of 1.08$\times$ and a peak speedup of 1.12$\times$. 
It should be noted that the efficiency of the tiling strategy is not fully reflected in this breakdown analysis since it is fundamental in all implementations of Figure \ref{fig.exp.Ablation}. Specifically, the former three settings adopt the same tiling strategy, which achieves best performance on \textit{SPIDER w. TC}. Meanwhile, the last setting only adjusts the tiling size for SpTC hardware.

\section{Related Work}

\textbf{Stencil Optimizations.} Traditional stencil optimizations on CPUs employ techniques such as vectorization\cite{henretty2013stencil, li2021reducing}, tiling\cite{wolfe1989more, song1999new, jin2001increasing, strzodka2010cache, bondhugula2016diamond}, and layout transformation\cite{henretty2011data, henretty2013stencil} to mitigate memory bottlenecks and enhance parallelism.
For GPU architectures, researchers develop comprehensive optimization strategies, incorporating techniques such as prefetching\cite{rawat2019optimizing}, loop unrolling\cite{gysi2021domain}, streaming\cite{rawat2018domain, zhao2019exploiting}, and more advanced tiling techniques specifically optimized for GPUs\cite{nguyen20103, holewinski2012high, grosser2013split,verdoolaege2013polyhedral, falch2014register, maruyama2014optimizing}. 
More recently, TCStencil\cite{TCStencil} demonstrates significant performance gains by transforming stencil computation as GEMM to leverage Tensor Cores.
Based on this approach, ConvStencil\cite{chen2024convstencil} and LoRAStencil\cite{zhang2024lorastencil} optimize the transformation process for more efficient execution. Furthermore, FlashFFTStencil\cite{han2025flashfftstencil} leverages FFT to enhance arithmetic intensity and optimizes the implementation on Tensor Cores with reduced memory access.

\textbf{Sparse ALU Assisted Acceleration.} Sparse ALU, first introduced by NVIDIA to accelerate sparse neural networks\cite{mishra2021accelerating}, has gained widespread adoption in deep learning. The straightforward approach to leverage Sparse ALUs utilizes vendor-provided libraries like cuSPARSELt\cite{cusparselt}. Beyond such libraries, nmSPARSE\cite{nmSparse} implements element-wise structured sparse kernels, and DFSS\cite{chen2023dynamic} develops a structured sparse attention kernel. 
However, conventional sparse ALU requires fixed 50\% sparsity. To address this, VENOM\cite{VENOM} and Samoyeds\cite{wu2025samoyeds} propose flexible sparse formats supporting configurable sparsity. Concurrently, Jigsaw\cite{zhang2025jigsaw} reorders unstructured sparse data into hardware-compatible structured sparsity.
Despite these advances, existing research utilizing Sparse ALUs remains primarily confined to GEMM kernels, leaving broader applicability to other computation patterns (e.g., stencil computation) unexplored.

\section{Conclusion}

We present SPIDER, a system to accelerate stencil computation using SpTCs, extending their application beyond deep learning. 
The transformation scheme in SPIDER is carefully designed to convert stencil computation into SpTC-compatible operations, enabling elegant implementation without introducing runtime overhead. 
Experimental results demonstrate significant performance advantage of SPIDER, achieving average speedups of 6.20$\times$ over vendor library cuDNN and 2.00$\times$ over Tensor Core-based state-of-the-art.

%% Bibliography
\bibliography{ref}

\end{document}